\documentclass[ letterpaper,10pt,conference ]{IEEEtran}
\usepackage{amsmath,amssymb,graphicx, epstopdf}
\usepackage[font={small,it}]{caption}
\usepackage{cite}
\usepackage{bm}
\newcommand*{\B}[1]{\ifmmode\bm{#1}\else\textbf{#1}\fi}
\DeclareMathOperator{\E}{\mathbb{E}}

\DeclareMathOperator{\PP}{\mathbb{P}}
\DeclareMathOperator{\NN}{\mathcal{N}}

\DeclareMathOperator{\Q}{\mathcal{Q}}

\begin{document}
%
\title{On Uplink User Capacity for Massive MIMO Cellular Networks}
%
%
%

\author{Anand Sivamalai and Jamie S. Evans

}
%
%

%


\maketitle

\begin{abstract}
Under the conditions where performance in a massive MIMO network is limited by pilot contamination, the reverse link signal-to-interference ratio (SIR) exhibits different distributions when using different pilot allocation schemes. By utilising different sets of orthogonal pilot sequences, as opposed to reused sequences amongst adjacent cells, the resulting SIR distribution is more favourable with respect to maximising the number of users on the network while maintaining a given quality of service (QoS) for all users. This paper provides a simple expression for uplink user capacity on such networks and presents uplink user capacity figures for both pilot allocation schemes for a selection of quality of service targets.

\end{abstract}

\section{Introduction}

Massive MIMO is a technology which is to play a significant role in tomorrow's 5G networks. From the currently available research, it is evident that multi-user massive MIMO technology when deployed on future cellular networks will bring significant gains to network sum rate \textit{data} capacity, whereas in previous cellular generation advancements, the attention was additionally on increasing the number of users the network could service, i.e. \textit{user} capacity. As we move into the ``Internet of Things" paradigm, the number of connected devices is estimated to be more than 25 billion by 2020 \cite{gartner}, it is clear we will need to turn our attention again to  ensuring \textit{user} capacity. While a large amount of research in massive MIMO systems focuses on cell sum rate capacity and spectral efficiency \cite{marzetta} \cite{Bjornson} \cite{hoydis} \cite{huh}, there has been little with a focus on user capacity \cite{shen} \cite{Akbar}. The recent work in \cite{Akbar} extends the single-cell user capacity expressions in \cite{shen} by presenting expressions for multi-cell user capacity and like \cite{shen} examines the \textit{downlink} user capacity while deriving optimal pilot sequences. In contrast to  \cite{Akbar}, this paper presents a comparison of two simple pilot allocation schemes, with a focus on \textit{uplink} user capacity and the gains that can be achieved through statistical multiplexing when admitting users on a network with the constraint of a minimum SIR and outage probability. 

The term ``massive" in massive MIMO, refers to the operating scenario where the BS has several orders of magnitude more antenna elements than users. It has been shown in \cite{marzetta}, that as the number of base station antennas $M$ becomes large, the effects of noise and uncorrelated intra and inter-cell interference disappear, and only inter-cell interference due to pilot contamination remains - the interference resulting from channel estimates which are contaminated from pilots of users in neighbouring cells. And hence a common tighter definition of ``massive" exists, where the system operates under such a regime where pilot contamination is the dominant impairment \cite{hoydis}. 

This paper starts by looking at the asymptotic result when the number of BS antennas is large \cite{marzetta}, and derives an analytical approximation for the distribution of the SIR, primarily based on a distance based path loss model and random user locations.  We then introduce the idea of statistical multiplexing with regards to user admission, and use our analytical SIR distribution to evaluate the number of users the network can support. 

Under the scenario where each BS uses the same set of orthogonal pilots, we expect one non-orthogonal interferer per adjacent cell. A way to combat such interference is to employ frequency reuse factors which are less than one, where adjacent cells utilise a different frequency resource, and consequently connected users of this cell are now orthogonal and no longer interferers. Such cellular frequency reuse patterns have a fixed granularity (Figure~\ref{figure:FreqReuse}) due to arranging the divided frequency resources in a regular fashion. Under a different pilot allocation scheme, where each cell uses an independent orthogonal set of pilots, every user in an adjacent cell is an interferer, albeit a fraction of the interference from a user under the former scheme. This provides a finer granularity to control the inter-cell interference before having to resort to reducing the frequency reuse - limiting the number of users will now effectively reduce the inter-cell interference, allowing it to come closer to the required QoS under user admission. On top of this, the reduced variability of the interferers in the context of statistical multiplexing under the second scheme makes it superior to the former scheme when maximising the number of users that can be admitted onto a network. The application of statistical multiplexing has been used extensively in the dimensioning of \textit{effective bandwidths} for users of Code Divison Multiple Access (CDMA) cellular networks. In \cite{evans2} several effective bandwidth models are presented, which are then used to develop capacity specifications and consequently call admission procedures for multi-cell CDMA networks with multiple classes. A very similar approach is employed in this paper to dimension an \textit{effective interference} for each user, which then enables the calculation of a user capacity.

\section{System Model and limiting SIR expressions}

\begin{figure} 
  \centerline{\includegraphics[scale = 0.6]{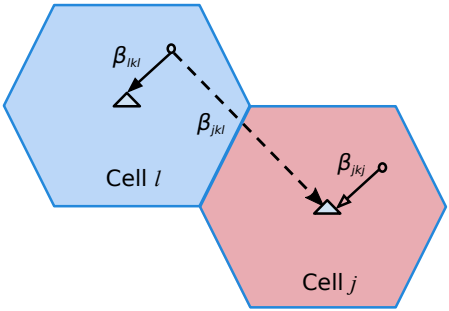}  }
    \caption{Uplink Pilot Contamination - Pilot reception from user of interest in cell j is contaminated with pilot transmission from interfering user in neighbouring cell l.}
    \label{figure:PilotContam}
\end{figure}

The problem of pilot contamination occurs when the base station receives indistinguishable uplink training pilot signals from both the intended user and other interfering users. Hence the estimated channel for the intended user is compromised, and subsequent data decoding suffers. Figure~\ref{figure:PilotContam} depicts such a scenario where the BS of cell $j$ receives uplink pilot transmissions from both its user and an interfering pilot transmission from a user in the neighbouring cell $l$.

We start with the result from \cite{marzetta}, which shows that in a cellular network with $L$ cells with $K$ users in each cell, as the number of antennas $M$ approaches infinity at BS $j$, all the effects of uncorrelated receiver noise and fast fading are eliminated, and we are left with pilot contamination as the dominant impairment. Consequently, as $M$ goes to infinity, the reverse link SIR at the BS for the $k$-th terminal, in the $j$-th cell, reduces to:

\begin{IEEEeqnarray}{rCl}  \label{eq:SIRlimit}
 SIR & = & \frac{\beta_{jkj}^2 }    { \sum_{l \neq j}\beta_{jkl}^2},
\end{IEEEeqnarray}
where $\beta$ represents the slow fading gain incorporating the effects of distance based path loss and shadow fading. Therefore, (\ref{eq:SIRlimit}) is the ratio of the slow fading gains of the $k$-th user in the $j$-th cell of interest, to the sum of slow fading gains to the target base station from all users in the network that employ the same pilot sequence. Interestingly, as highlighted in \cite{marzetta} the SIR is proportional to a ratio of the \textit{squares} of the slow gains, which is a result of the MRC processing and the interference being the dominant impairment.

Note, due to the elimination of noise terms in (\ref{eq:SIRlimit}), $SINR$ is written throughout as $SIR$ for clarity. Given the fact that the noise and fast fading terms are no longer present, the simplicity of (\ref{eq:SIRlimit}) allows us to analyse its distribution further, which can then be used as a basis for evaluating the admissibility of a user on a network. 

We acknowledge that the convergence to the asymptotic behaviour of (\ref{eq:SIRlimit}) is slow, typically requiring an impractical number of BS antenans even in the absence of noise, at least $M = 10^5$ were required in our simulations to come within 15\% of the mean of the large $M$ SIR, where similar observations are noted in \cite{Bjornson}. However the large $M$ result serves as a means to compare the two pilot allocation schemes analytically, where later, the finite $M$ Monte Carlo simulation results confirm the differences of the two schemes. Some very recent work by \cite{Bai} presents some accurate approximations to the SIR distribution for the reverse link under a similar system model for finite $M$, but with the main difference that the users and BSs are distributed as Poisson point processes. Our approach follows the more traditional fixed hexagonal grid of BSs and uniformly distributed users, and allows for a simple explicit expression for user capacity from the key system parameters.

\subsection{Uplink Power Control}

In order to improve inter-cell interference, uplink power control (ULPC) can be utilised on the reverse link. We incorporate a modest open loop scheme which does not attempt to compensate for fast fading.  We assume perfect path loss estimation at the terminal, with each terminal transmitting with a power which is the inverse of the path loss to its own base station. The terminal transmits the pilots and the corresponding data at the same power, and (\ref{eq:SIRlimit}) becomes:

\begin{IEEEeqnarray}{rCl}  
 SIR & = & \frac{\:(\frac{\beta_{jkj}}{\beta_{jkj}})^2 }    { \sum_{l \neq j}(\:\frac{\beta_{jkl}}{\beta_{lkl}})^2      }\IEEEnonumber
\\                            \label{eq:SIRlimit_ulpc1}
& = & \frac{1}    {  \sum_{l \neq j}(\frac{\beta_{jkl}}{\beta_{lkl}})^2      }.  
  \end{IEEEeqnarray}

\subsection{Pilot allocation}

The result in (\ref{eq:SIRlimit_ulpc1}) represents a pilot allocation where each cell in the network re-uses the same set of orthogonal pilots. As a result, when $M$ is large, there is no intra-cell interference in the $j$-th cell, and there is one interfering user in every other cell in the network, resulting in a denominator that is a sum of $L-1$ terms. 

An alternative pilot allocation, which will be used to compare against the aforementioned re-used pilot scheme in this paper, is to allocate different orthogonal pilot sets in all cells of the network. Such an approach still leaves us with no intra-cell interference but now \textit{every} inter-cell user in the network is an interferer, since all these users have used a pilot sequence which is no longer orthogonal to the pilot of the user of interest. This results in an interference which is the sum of $K \times(L-1)$ terms. In our SIR expression, this non-orthogonality is represented by $\phi_{kl}$ which multiplies each of the interference terms, and is equal to the square of the inner product of the pilot sequence of the $j$-th user, with the pilot sequence of the interfering user. In contrast, under the reused pilot set scheme, $\phi_{kl} = 1$ for only one user in the $l$-th cell and $\phi_{kl} = 0$ for all remaining users of the $l$-th cell. (This approach models $\phi_{kl}$ as a random quantity as opposed to (26) in \cite{marzetta}, where only the expected value of $\phi_{kl}$ is used instead in the limit expression). As a result, (\ref{eq:SIRlimit_ulpc1}) becomes :

\begin{IEEEeqnarray}{rCl}  \label{eq:SIRlimit_ulpc2}
 SIR = \frac{1}    { \sum_{l \neq j}\sum_{k=1}^{K}\phi_{kl}(\:\frac{\beta_{jkl}}{\beta_{lkl}})^2      }.
\end{IEEEeqnarray}

\section{Analysis of the SIR distribution}

In our analysis of (\ref{eq:SIRlimit_ulpc2}), we model the slow fading gain by distance based path loss alone, where $\beta_{jkl} = r_{jkl}^{-\gamma}$. The users are uniformly distributed over the cell area, where the random quantity $r_{jkl}$, is the distance between the $k$-th user in the $l$-th cell and BS $j$. The constant $\gamma$ is the path loss exponent. Including the effects of log-normal shadowing in the slow fading gain of our analysis of the SIR expression is non-trivial, and is further discussed in Section \ref{sec:LogNormalFading2}.

Therefore, we begin by defining,
\begin{IEEEeqnarray}{rCl}  \label{eq:ykl}
y_{kl} =  \phi_{kl} \left(\frac {r_{lkl}}{r_{jkl}} \right)^{2 \gamma},
  \end{IEEEeqnarray}

so that the SIR in (\ref{eq:SIRlimit_ulpc2}) can be expressed as:
\begin{IEEEeqnarray}{rCl}  \label{eq:SirXi}
 SIR = \frac{1}  { \sum_{l \neq j}\sum_{k=1}^{K} y_{kl}}.
  \end{IEEEeqnarray}
  
Since we are interested in determining the distribution of the SIR, it is clear that we need to determine the distribution of the sum of the random variables in the denominator. The random quantity $y_{kl}$ is a function of three different random variables ${r_{lkl}}$ , ${r_{jkl}}$ and inner product of the pilot sequences $\phi_{kl}$. Therefore the nature of the SIR expression is still quite complex. Rather than attempting to find the exact distribution we will approximate the sum of $y_{kl}$ random variables by a Gaussian random variable, based on the central limit theorem. In order to make this approximation, we need the mean and variance of the quantity $y_{kl}$.

\begin{figure} 
  \centerline{\includegraphics[scale = 0.6]{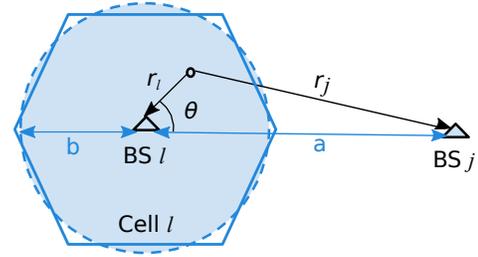}  }
    \caption{Circular cell approximation with user in cell $l$ interfering with reverse link reception at BS $j$}
    \label{figure:CircleApprox}
\end{figure}

In the interests of readability for the remaining part of this section, we can drop the $kl$ subscript by just considering the two BS scenario shown in Figure~\ref{figure:CircleApprox}, with our BS of interest (BS $j$) and an arbitrary interferer in an adjacent interfering cell (cell $l$). Note that we have approximated the hexagonal cell by a circle as will be explained in more detail shortly.

Due to the statistical independence between the pilot sequences and the remaining random quantities, the mean can be written as $\E\left[ y \right] = \E\left[ \phi \right] \E\left[ x\right]$, where the random variable $x = \left( {r_{l}}/{r_{j}}\right)^{2 \gamma}$. The $\E\left[ x\right]$ can be determined analytically as follows.

Assuming users which are uniformly distributed over each cell of our cellular network, the probability distribution functions (PDFs) of the two random variables $r_l$ (the distance of the interferer to their own BS) and $r_{j}$ (the distance of the interferer to the base station of interest) are dependent on the geometry of the individual cells and the layout of the cells within the network. These two random quantities are clearly not independent and therefore, computing a joint PDF is difficult.

An alternative approach, which is also used in \cite{evans1}, is to write
\begin{IEEEeqnarray}{rCl}  \label{eq:Interference}
x = \left(\frac {r_l}{r_j} \right)^{2 \gamma} = v(r_l,\theta) =  ( \frac{r_l^2} {{r_l ^2 + a^2 - 2 a r_l  \cos \theta}})^{\gamma}.
 \IEEEnonumber
  \end{IEEEeqnarray}
If we now approximate the hexagon shaped cell with a circular cell, we achieve two important points. The circle cell approximation results in very simple PDFs, where $f_{Rl}(r_l) = 2r_l/{b}^2$ for $0 \leq r_l < b$ and $f_\Theta(\theta) = 1/\pi$ for $0 \leq \theta < \pi$. Secondly, we have statistical independence between $r_l$ and $\theta$, and can write 
\begin{IEEEeqnarray}{rCl}  \label{eq:ExpX}
\E\left[ x \right] = \int_0^\pi \int_0^b v(r_l,\theta)f_{Rl}(r_l)f_\Theta(\theta) dr_l d\theta 
 \\
 \label{eq:VarX}
\sigma_x^2 = \int_0^\pi \int_0^b (v(r_l,\theta) - \E\left[ x \right])^2 f_{Rl}(r_l)f_\Theta(\theta) dr_l d\theta 
  \end{IEEEeqnarray}

Given (\ref{eq:ExpX}) and (\ref{eq:VarX}) we are able to make the Gaussian approximation in (\ref{eq:SirXi}) under the pilot allocation scheme where the same orthogonal set is reused in every cell, (since effectively $\phi$ is a constant). If we now consider the scheme where each cell uses a different orthogonal set, we will have to derive the mean $\mu_{y}$ and variance $\sigma_{y}^2$ where the random nature of  $\phi$ is treated. The pilots sequences from the orthogonal set are assigned randomly to the $K$ users of the cell, therefore $\phi$ is statistically independent from $x$, (the user locations). The pilot sequences $\psi$ are the coloumns from a $K \times K$ unitary matrix, distributed uniformly according to the Haar measure. Consequently, the random quantity $\phi = |\left\langle \psi_{j}, \psi_{l}\right\rangle|^2 $, has an expectation of $1/K$ and a variance of $1/K^2$, and the mean is given by:
\begin{IEEEeqnarray}{rCl}  \label{eq:mean_y}
 \mu_{y} & = & \E\left[ \phi x \right] =   \E\left[ \phi \right] \mu_{x} 
 =   \mu_{x}/K .
  \end{IEEEeqnarray}

In order to derive the variance  $\sigma_{y}^2$, a series of simple substitutions can be made, resulting in:

\begin{IEEEeqnarray}{rCl} \label{eq:var_y}
 \sigma_{y}^2 & = &  \frac{1}{K^2} ( 2 \sigma_{x}^2 + \mu_{x}^2).
  \end{IEEEeqnarray}

%

\section{User capacity and effective Interference}

One of the objectives of any user admission policy is to ensure that admitting a new user to the network will still guarantee a certain QoS for the existing users of a network. In the context of our massive MIMO network, whose performance is interference limited, such an admission policy would attempt to predict if admitting a new user would add an acceptable level of interference to all users of the network. 

As seen in the previous section, the SIR expression in (\ref{eq:SIRlimit_ulpc2}) is random in nature and hence not easy to predict accurately. Just as \cite{evans2} describes an \textit{effective bandwidth}, we introduce the notion of an \textit{effective interference}, $y_{E}$ for each user. We assign an effective interference for each user which is somewhere in between the mean interference and the maximum interference. Assigning an effective interference for each user which is equal to the maximum interference results in a very conservative admission policy which does not benefit at all from statistical multiplexing. Assigning an effective interference equal to the mean interference implies the unrealistic scenario of an infinite number of users on the network, where the sample mean is infact the true mean.

A user admission policy would ensure that every user of the network experiences a quality of service, governed by a minimum SIR, $S$ and an outage probability $\alpha$. Using the Gaussian approximation for the denominator of (\ref{eq:SIRlimit_ulpc2}), this condition can be written as:
\begin{IEEEeqnarray}{rCl}  
 \PP (\frac{1} { \NN (\mu, \sigma)}  >  S) \geq 1 - \alpha,  \IEEEnonumber 
   \end{IEEEeqnarray}
which is equivalent to the condition,
 \begin{IEEEeqnarray}{rCl}  
     \frac {1/S - \mu}{\sigma} \geq \Q^{-1}(\alpha),  \label{eq:qos}
  \end{IEEEeqnarray}
where $\Q(x) = 1 - \Phi(x)$, and $\Phi(x)$ is the CDF function for $\NN (0,1)$. 

The central limit theorem allows us to approximate the sum of interference from the $n$ users, which are \textit{i.i.d} \footnote{In this approximation, we consider the weak correlation in the interference terms induced by the unitary property of the pilot sequence matrix negligible.}, by a Gaussian random variable with $\mu = n \mu_{y}$ and $\sigma^2 = n \sigma^2_y$. However, depending on the network topology, the surrounding cells containing the interferers could be at different distances, resulting in interference with different means and variances. The Gaussian approximation can be extended to approximate the sum of these $T$ different types of interferers, where the total interference is now approximated by a sum of $T$ normal distributions, which is itself also a normal distribution, with $\mu = \sum_{t=1}^T n_t \mu_{y_t}$ and $\sigma^2 =  \sum_{t=1}^T n_t \sigma^2_{y_t}$, resulting in:
\begin{IEEEeqnarray}{rCl}
          \frac {1/S - \sum_{t = 1}^T n_t\mu_{y_t}}{\sqrt{\sum_{t = 1}^T n_t \sigma_{y_t}^2}} \geq \Q^{-1}(\alpha).  \label{eq:qos2}   
  \end{IEEEeqnarray}

If we consider the traditional cellular layouts as shown in Figure~\ref{figure:FreqReuse}, we initially only need to consider the Tier 1 interferers ($t \leq T = 1$) - the set of interferers (marked in dark blue for different frequency reuse factors) which are closest to our user/cell of interest (marked in red). This is because under our current model, the interference contribution from users in the next tiers of interfering cells ($1 < t \leq T$), is negligible. For instance, given a frequency reuse factor of 1 (with $x_{t=n}$ denoting the random interference at tier $n$), we have $\E\left[ x_{t=1} \right] \simeq 500 \E\left[ x_{t=2} \right]$ , and $\mathrm{var} \left[ x_{t=1} \right] \simeq 10^6 \mathrm{var}\left[ x_{t=2} \right]$. Under some extensions to our model, the outer tier interferers do become relevant as discussed in Section \ref{sec:MultiTier}.

Using (\ref{eq:qos2}) with $T = 1$, and for a given $\alpha$ and $S$, we can solve for the maximum number of interferers $n_{\max}$, where $n_{\max} = \lfloor n_1 \rfloor $. As a result, each of these interferers contributes an \textit{effective} interference $y_E$, given in [3] as :
\begin{IEEEeqnarray}{lrCl}   \label{eq:y_E}
y_E = \mu_1 / (1 + \frac{2}{z}(1 - \sqrt{1+z})), 
 \end{IEEEeqnarray}
where
\begin{IEEEeqnarray}{lrCl}  
 z = \frac{4 \mu_1 }{ (\Q^{-1}(\alpha))^2 \sigma_1^2 S}.     \IEEEnonumber
\end{IEEEeqnarray}

Therefore, from (\ref{eq:y_E}), the maximum number of allowed interferers $n_{\max}$ across all Tier 1 interfering cells, in our power controlled system is given by: 
\begin{IEEEeqnarray}{lrCl}   \label{eq:n_max} 
n_{\max} \leq \frac {1} {y_E S}. 
\end{IEEEeqnarray}

\section{Numerical Results} \label{sec:NumericalResults}
\subsection{Scenario} \label{sec:Scenario}

Our system is modelled around the LTE reverse-link in \cite{marzetta}. Consequently we consider the possible frequency reuse factors $w \in\lbrace1,3,7\rbrace$, assume a coherence bandwidth of 14 subcarriers, and our coherence time is divided into 7 symbols, 3 of which are used for uplink training. As a consequence, our pilot sequence is of length $3 \times 14 = 42$, and can therefore support a maximum of $K = 42$ terminals. Each cell has a radius $a = 1600$ metres and with cell-hole radius $a_{h} = 100$ metres. A path loss exponent of 4 is used.

We wish to model a simple non-cooperative user admission policy, and therefore impose the requirement that each BS can decide to admit a user autonomously, (i.e. BSs do not require information from other BSs for user admission) which in turn implies that the maximum number of users connected to a BS is the same for all BSs of the network.

\begin{figure} 
  \centerline{\includegraphics[scale = 0.5]{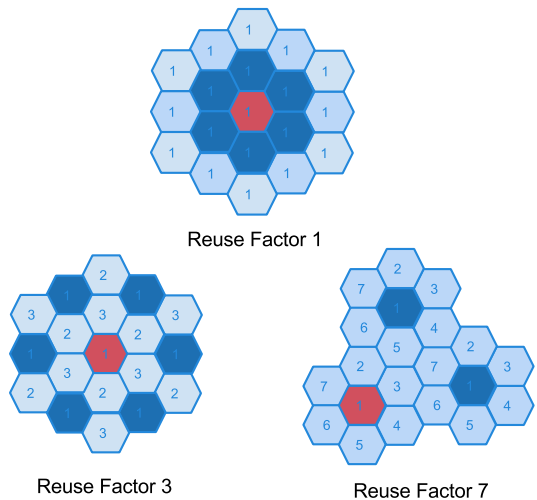}  }
    \caption{Non-orthogonal Tier 1 interferers (dark blue) under different frequency reuse factors $w$.}
    \label{figure:FreqReuse}
\end{figure}

In order to meet higher QoS requirements, we employ frequency reuse factors $w$ using traditional cellular frequency reuse patterns to reduce the interference experienced by the user of interest. By utilising different frequency resources in adjacent cells, all users in adjacent cells no longer interfere with the user of interest (located in the cell marked in red of Figure~\ref{figure:FreqReuse}), and non-orthogonal interferers are essentially moved further away (who are located in the cells marked in darker blue). Of course this reduction of interference to the user of interest comes at the cost of a reduction (by the frequency reuse factor) to the maximum number of users that can be supported by each cell, since the available frequency resources within the cell to train its users has been reduced.

Figure~\ref{figure:FreqReuse} shows the different frequency reuse factors $w$, with the cell of interest in red, and the cells which have interfering users in darker blue. The frequency resource utilised by the cell is indicated by the number in the cell.

\subsection{M in the Limit} \label{sec:Mlimit}

As highlighted earlier we consider only Tier 1 interfering cells, and from (\ref{eq:y_E}), the effective interference $y_E$ is computed for a given minimum SIR and outage probability. Using (\ref{eq:n_max}), the total maximum number of interferers across all Tier 1 interfering cells $n_{\max}$ is given, and hence the \textit{unconstrained} maximum number of admissible users \textit{per} cell is given by, $k_{u} = n_{\max}/6 $. However, given that the number of users a cell can support is constrained by the length of the pilot sequence and our simple non-cooperative admission policy, the number of users per cell which we can support in practise is given by $k_{\max}$. 

  \begin{figure} 
  \centerline{\includegraphics[scale = 0.35]{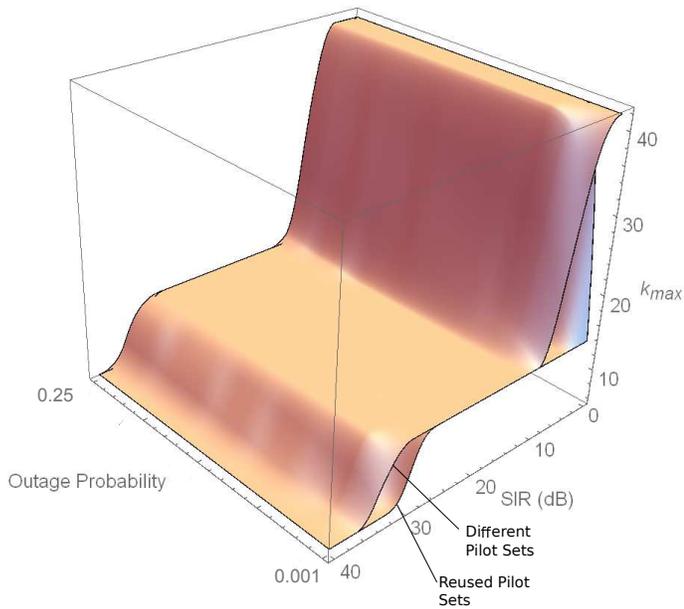}  }
    \caption{Maximum number of users per cell, $k_{\max}$ vs. QoS}
\label{figure:InfiniteM_3D}
\end{figure} 

Figure~\ref{figure:InfiniteM_3D} shows clearly that for the region of relevant outage and SIR requirements, a pilot allocation scheme which employs different pilot sets, as opposed to reused sets, is superior with regards to $k_{\max}$.

\begin{figure} 
  \centerline{\includegraphics[scale = 1.0]{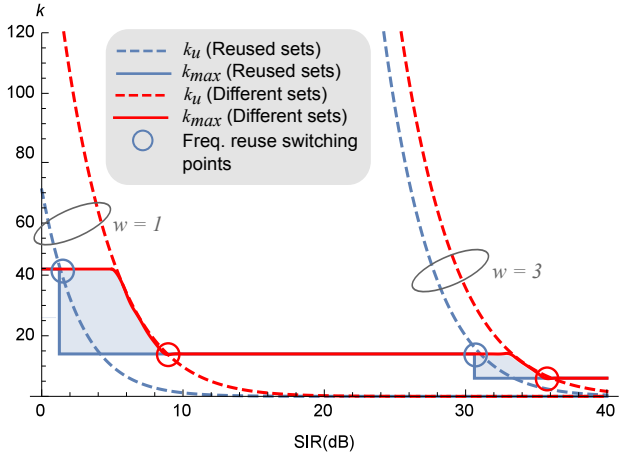}  }
    \caption{Maximum number of users per cell, $k_{\max}$ and Unconstrained maximum number of users $k_{u}$ vs. SIR under fixed outage probability of 0.05}
\label{figure:InfiniteM_SIR_vs_k}
\end{figure}

In order to examine the differences further, Figure~\ref{figure:InfiniteM_SIR_vs_k} shows a cross-section of Figure~\ref{figure:InfiniteM_3D}, with both $k_{u}$ and $k_{\max}$ plotted for a fixed outage probability of 0.05. Firstly, for both $w=1$ and $w=3$, it is noteworthy that $k_{u}$ is significantly higher under the relevant SIR range when different pilot sets are used.  

Under low SIR requirements, for both schemes, the number of users in the cell is restricted by the resources allocated to uplink training - as $k_{u}$ shows we would be able to support many more users at this QoS based on their interference contribution, but can only support the 42 based on our pilot length. Therefore the maximum number of users that can be supported by the BS, $k_{\max}$, is the same, regardless of the pilot allocation scheme. However, to meet an SIR requirement which is greater than 1dB when pilot sets are reused, we are forced to employ $w=3$, indicated by the ``switching point" circled on Figure~\ref{figure:InfiniteM_SIR_vs_k}, while when using different pilot sets we are able to continue using $w=1$. Moving to $w=3$ reduces the effective interference $y_E$ by a factor of 500 since the interferers are now positioned in cells further away, and as can be seen from the steep slope of the $k_{u}$ curves for both schemes under $w=3$, we would be able to support many users (for all SIRs up to 26dB). However, frequency resources for uplink training have consequently been reduced by a factor of 3, and in practice we can only train a maximum of $42/3 = 14$ users. 

Under the different pilot set scheme, $k_{\max}$ only starts to decline for SIRs greater than 5dB, where we are able to support the increasing SIR requirements by simply admitting fewer users in the cell. For SIRs greater than 9dB we would have to admit less than 14 users to meet this requirement, and therefore it makes more sense that we employ $w=3$ (as indicated by the switching point, where the red $k_{u}$ curve falls under the $k_{\max}$ line). It is the region between these switching points, indicated by the shaded blue in Figure~\ref{figure:InfiniteM_SIR_vs_k}, which is the region of gain when using the different pilot sets scheme. Both schemes now support the same number of users up until an SIR around 30dB, where the reused pilot set scheme must again switch to $w=7$, while the different pilot set scheme only switches at an SIR of 35dB, and hence can support more users.

As can be seen clearly in Figure~\ref{figure:InfiniteM_3D}, the outage probability component of the QoS has less of an effect on $k_{\max}$, when compared to changes in the SIR.  The shaded regions in Figure~\ref{figure:InfiniteM_SIR_vs_k} clearly show for the given QoS, the scale and region where significant gains in user capacity can be realised by utilising different orthogonal pilot sets amongst cells.

\begin{figure} 
  \centerline{\includegraphics[scale = 0.58]{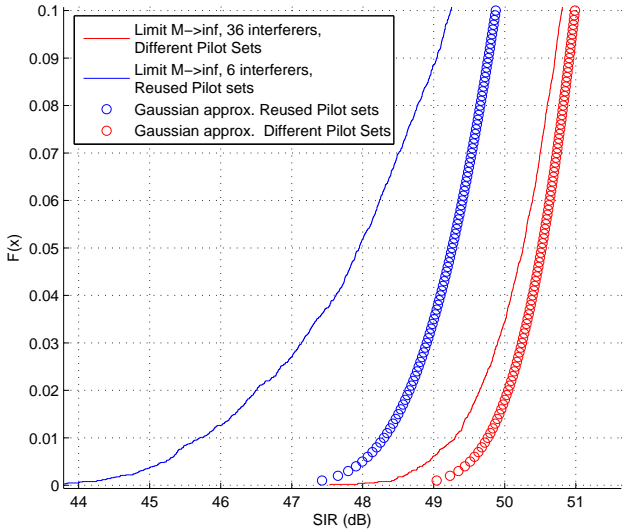}  }
    \caption{SIR CDF under Freq. Reuse factor = 7 - (Worst-case) Gaussian Approximation vs. Simulation, under large M}
\label{figure:GaussApprox}
\end{figure}

We now briefly look at the \textit{worst} case accuracy of the Gaussian approximation under large $M$ used to generate the results shown in Figure \ref{figure:InfiniteM_SIR_vs_k}. As posed by the central limit theorem, we expect that as the numbers of interferers increases, the distribution of interference closer approaches that of a normal distribution, and the accuracy of our estimation improves. Given our restriction of a simple user admission policy and standard cellular frequency reuse patterns, our \textit{worst} case approximation would be under a frequency reuse factor of 7. Under this level of frequency reuse we could support a maximum of 6 users in each of the interfering Tier 1 cells. Figure \ref{figure:GaussApprox} shows a comparison between the Gaussian approximation (in combination with our circular cell approximation) and a Monte Carlo simulation, for reverse link SIR distributions for both pilot allocation schemes under frequency reuse factor 7. When specifying meaningful QoS, we typically deal with outages of 0.1 or less and therefore the main area of interest is in the lower tail of the distributions. As expected, the approximation of the different pilot set scheme follows more closely the results of the simulation, a consequence of being the sum of 36 interferers from these 6 interfering cells, as opposed to 6 interferers in the reused pilots case. Importantly, our approximated user capacity gain from the different pilot set allocation is conservative, since around this region, the approximation returns results significantly more optimistic in the reused pilot sets case compared to when different pilot sets are used.

\subsection{Finite M} \label{sec:FiniteMResults}

As discussed previously, the asymptotic behaviour of the reverse link SIR as $M$ grows large results in the noise and fast fading becoming negligible due to the sum of the un-correlated cross terms approaching zero. Consequently, we are able to derive the approximation for the reverse link SIR analytically. In the case of finite $M$, all such cross terms are present, and consequently the expression for the reverse link SIR at the BS contains many more random variables. 

Given the complexity of the SIR expression under finite $M$, we present results from a Monte Carlo simulation in Table \ref{table:MaxNumUsersQoSFiniteM} for the maximum number of users that can be admitted per cell for a given arbitrary set of QoSs. We present some results here to show that significant real world gains are also expected in terms of user capacity, where the values presented here are only to serve as an indication. Due to the presence of \textit{all} intra and inter-cell interference terms, we expect estimates of user capacity which are less than our previous analysis with $M$ in the limit. Given recent measurements from real world massive MIMO experiments, antenna arrays of 128 elements \cite{gao1} \cite{gao2} have been used and we select an $M$ of similar order, where $M = 500$. From Table \ref{table:MaxNumUsersQoSFiniteM}, it can be seen that as soon as we increase the frequency reuse factor $w$ to meet the given QoS, the interference levels fall well below the requirement, and we are then simply limited by the reduced available uplink training resources.

\begin{table}[!t]
\renewcommand{\arraystretch}{1.3}
\caption{Maximum Number of Admissible Users per cell for Different QoS, M = 500, MRC detector}
\label{table:MaxNumUsersQoSFiniteM}
\centering
\begin{tabular}{lllll}
\hline
QoS                                                             & 
\begin{tabular}[c]{@{}l@{}}Low \\ 0db/0.01\end{tabular} & 
\begin{tabular}[c]{@{}l@{}}Medium \\ 10db/0.05\end{tabular} & 
\begin{tabular}[c]{@{}l@{}}High \\ 25db/0.05\end{tabular} & 
\begin{tabular}[c]{@{}l@{}}Very High\\ 30db/0.005\end{tabular} \\ \hline
\begin{tabular}[c]{@{}l@{}}Reused \\ Pilot Sets\end{tabular}    & 14                                                     & 14                                                          & 6                                                         & 6                                                               \\ \hline
\begin{tabular}[c]{@{}l@{}}Different \\ Pilot Sets\end{tabular} & 42                                                     & 14                                                          & 14                                                         & 6                                                              \\ \hline
\end{tabular}
\end{table}

\section{Extensions}

\subsection{Generalised Cooperative Admission Policy} \label{sec:CooperativeAdmission}

Since all Tier 1 interferers in our system are equivalent, a \textit{cooperative} user admission policy could manage user admission across the set of cells $T_1^{(j)}$ which are considered Tier 1 interfering cells to cell $j$, by ensuring only the \textit{sum} of all users in these cells is less than the combined amount $n_{\max}$. Under such a policy, this constraint would have to be satisfied for every cell in the network.   i.e.:
 \begin{IEEEeqnarray}{lrCl}   \label{eq:GeneralisedCooperative} 
\sum_{l \in T_1^{(j)}} k_{l} \leq  n_{\max}, \forall j \in L.
\end{IEEEeqnarray}

Furthermore, the upper limit of users within the cell is still restricted by the length of the pilot sequence, i.e. $k_{l} \leq K/w$.
Such a policy would increase the probability of a randomly placed new user being admitted at the expense of cooperation between the BSs of the network.
\subsection{Log-normal shadowing and BS selection}\label{sec:LogNormalFading2}

\subsubsection{BS selection} \label{sec:BSselection}

The path loss due to shadowing can be modelled by a log-normal random variable $z$, where typically for macrocell shadowing we have $ \NN (0,8) = 10\log_{10}z$. When this is to be modelled by our slow fading gain, we have $\beta_{jkl} = z_{jkl}\:r_{jkl}^{-\gamma}$, and (\ref{eq:ykl}) becomes :

\begin{IEEEeqnarray}{rCl}  \label{eq:ykl2}
y_{kl} =  \phi_{kl} \left(\frac {z_{jkl}}{z_{lkl}}\right)^2 \left(\frac {r_{lkl}}{r_{jkl}} \right)^{2 \gamma}.
  \end{IEEEeqnarray}

Due to the large variance of the random quantity $\left(z_{jkl}/z_{lkl}\right)^2$ in (\ref{eq:ykl2}), it can be expected that a given user may not necessarily experience the best channel to the closest BS, or even to the immediately surrounding BSs. Of course we would expect that a \textit{realistic} user admission policy would assign the user to the BS to which it experiences the best channel gain. As a consequence, for the $k$-th user in the $l$-th cell to be considered as an interferer for a user in the $j$-th cell, the condition $\beta_{jkl} \geq \beta_{lkl}$ must hold. From this inequality, we then have the constraint on the interference:

\begin{IEEEeqnarray}{rCl}  \label{eq:InterferenceConstraint}
\left(\frac {z_{jkl}}{z_{lkl}}\right)^2 \left(\frac {r_{lkl}}{r_{jkl}} \right)^{2\gamma} < 1,
  \end{IEEEeqnarray}

where the $l$-th cell may or may not be the closest cell to the user of interest. As a result of (\ref{eq:InterferenceConstraint}), a true statistical independence between these ratios no longer exists. In order to provide results for $k_{\max}$ as in Section \ref{sec:NumericalResults} for this model, (\ref{eq:InterferenceConstraint}) first needs to be properly incorporated in the analytical expression for SIR. 

\subsubsection{Multi Tier user admission} \label{sec:MultiTier}

When realistically modelling log-normal shadow fading in our system, results from Monte Carlo simulation show we must consider Tier 2 interferers since they are no longer negligible.

\subsection{Downlink User Capacity} \label{sec:DLcapacity}

In order to provide the inputs to a comprehensive user admission policy, downlink capacity also needs to be considered. As pointed out in \cite{marzetta}, the interferers in the SIR forward link expression are no longer strictly \textit{i.i.d}, and consequently we require an alternate approach to analyse downlink capacity. The expressions presented in \cite{Akbar} are based on large $M$ SINR results, and provide an alternative approach into dimensioning the downlink user capacity.

\section{Conclusion}

We derived a reverse link SIR expression for our system model, which was then used as a basis to derive an explicit expression for the maximum number of users which can be admitted to a cell in a multi-cell massive MIMO network, when $M$ is large. 

For both large and finite $M$, it has been shown that using different orthogonal pilot sets in each cell, as compared to reusing pilot sets amongst all cells, allows us to admit the same or significantly more users while upholding a given QoS for all users of the network.


\bibliographystyle{IEEEtran}
\bibliography{MIMO,IEEEabrv}

\end{document}